\documentclass[namedreferences]{solarphysics}

\usepackage[hyperref,optionalrh,showbiblabels]{spr-sola-addons} 
\usepackage{graphicx}        
\usepackage{color}           
\usepackage{breakurl}        

\chardef\us=`\_

\begin{document}

\begin{article}
\begin{opening}

\title{Rearrangements of Open Magnetic Flux and Formation of Polar Coronal Holes in Cycle 24}

\author[addressref={aff1},email={e-mail.golubeva@iszf.irk.ru}]{\inits{E.M.}\fnm{E.M.}~\lnm{Golubeva}\orcid{https://orcid.org/0000-0003-3040-8499}}
\author[addressref=aff1,corref,email={e-mail.avm@iszf.irk.ru}]{\inits{A.V.}\fnm{A.V.}~\lnm{Mordvinov}\orcid{https://orcid.org/0000-0001-8154-3899}}
\address[id=aff1]{Institute of Solar-Terrestrial Physics of  Siberian Branch of  Russian Academy of Sciences, Lermontov st., 126a, Irkutsk 664033, Russia }

\runningauthor{E.M.~Golubeva, A.V.~Mordvinov}
\runningtitle{Open Magnetic Flux and Polar Coronal Holes in Cycle 24} 

\begin{abstract}
A method of synoptic map assimilation has been developed to study global rearrangements of open magnetic flux and formation of polar coronal holes (PCHs) in the current cycle. 
The analysis reveals ensembles of coronal holes (ECHs) which appear within unipolar magnetic regions associated with decaying activity complexes (ACs). 
The cause-effect relations between them explain asynchronous PCH formation observed at the northern and southern hemispheres of the Sun. 
Thus, the decay of large ACs, that were observed in 2014, led to formation of an extensive ECH, which then became the south PCH in mid-2015. 
Intricate structure of magnetic fields in the northern polar zone has impeded the north PCH formation, despite the fact that the dominant polarity at the North Pole reversed two years earlier than at the South Pole. 
The north PCH formed only by mid-2016 as the result of gradual merger of high-latitudinal ECHs associated with several decaying ACs.
\end{abstract}

\keywords{Magnetic fields, Photosphere; Coronal Holes; Magnetic Reconnection, Observational; Solar Cycle, Observations}

\end{opening}

\section{Introduction}
     \label{S1-Intro} 
Spectro-polarimetric measurements demonstrated that the Sun$'$s polar fields are highly structured \citep{Tsuneta08}. Small-scale vertical fields of kilogauss strength coexist with nearly horizontal fields \citep{Harvey07}. However, the averaged high-resolution polar-fields are about 5 G. These weak open fields form a dipole-like structure. The polar caps are mostly covered by magnetic fields of dominant polarities. Polar coronal holes (PCHs) overlie the caps during most of a solar cycle. 
 
The Sun's open magnetic fields extend outward, forming the heliosphere. Most of the total open flux originates in coronal holes (CHs). Properties of high-speed solar wind streams depend on the geometry of open field lines \citep{Wang90}. 
The open flux footprints appear also over large-scale unipolar regions which are usually associated with large decaying active regions (ARs) and with magnetic plages \citep{Schrijver03}.  As an 11/22-year cycle progresses, the open and closed magnetic fields reconnect \citep{Antiochos07}. These interactions change the distribution of open flux over the solar surface \citep{Mackay02, Wang04}. 

The reversals of polar fields lead to global rearrangements of the Sun$'$s open magnetic-fields. During these periods, PCHs disappear. 
The causal relationship between the Sun$'$s polar magnetism and ARs is of crucial importance to understand cyclic changes in large-scale magnetic fields \citep{Petrie15,Golubeva16}. The field rearrangements change space weather conditions near Earth and over the whole heliosphere \citep{Sheeley15}.

\cite{Harvey02} have shown a systematic movement of mid-latitudinal CHs to high latitudes at the ascending phase of a solar cycle. As unipolar magnetic regions (UMRs) of opposite polarities approach the poles, existing PCHs decay, and their open fluxes redistribute over the solar surface. This rearrangement is completed with magnetic polarity changes in the polar caps and with formation of new PCHs \citep{Wang09,Petrie15}.

Ensembles of coronal holes (ECHs) usually form in UMRs associated with the decaying activity complexes (ACs). Long-lived ECHs play a key role in cyclic reorganization of open magnetic flux \citep{Golubeva16}. Evolution of ECHs is determined by interchange reconnection of open and closed magnetic fields, as well as by their interaction with other ECHs and PCHs. As UMRs approach the poles, ECHs appear at increasingly higher latitudes. Gradual accumulation and merger of ECHs in the polar regions lead to formation of PCHs.

In general, these observational facts agree with the flux transport dynamo \citep{Choudhuri95, Kitchatinov11, Karak14} and the numerical models which describe evolution of photospheric magnetic-fields \citep{DeVore85, Sheeley85, Sheeley87, Wang89, Mackay02, Baumann04, Jiang14,Jiang_14}. 
However, the formation of PCHs and the general mechanism of CH evolution are still not well-studied. In addition, the development of solar activity shows different peculiarities from cycle to cycle. Moreover, unusual development of solar activity in the current cycle resulted in asynchronous formation of the northern and southern PCHs \citep{Golubeva16,Lowder17}.

We study PCH formation and the rearrangement of open magnetic flux over the solar surface in the current cycle, analyzing synoptic maps of magnetic fields and CHs. Such an approach is useful to understand the evolution of CHs and the general mechanism of open flux rearrangement. We consider the global rearrangement of magnetic fields in relation to emergent magnetic fluxes, which mainly determined the current cycle peculiarities.

\section{Data}
\label{S2-Dt}
This study is based on the joint analysis of synoptic maps which represent magnetic fields and CHs for Carrington Rotations (CRs) 1910\,--\,2190 (1996.42-2017.40 yrs). We use the maps of radial component of magnetic field from the \textit{National Solar Observatory/Kitt Peak Vacuum Telescope} (NSO/KPVT), the \textit{Synoptic Optical Long-term Investigations of the Sun/Vector Spectro-magnetograph} (SOLIS/VSM), and the \textit{National Solar Observatory/Global Oscillation Network Group} (NSO/GONG). Corresponding synoptic maps of CHs from NSO/KPVT, the ESA-NASA \textit{Solar and Heliospheric Observatory/Extreme ultraviolet Imaging Telescope} (SoHO/EIT), and the \textit{Solar Dynamics Observatory/Atmospheric Imaging Assembly} (SDO/AIA) involve CRs 1910\,--\,2163 (1996.42\,--\,2015.39 yrs). To plot the missing maps for CRs 2164\,--\,2190 (2015.39\,--\,2017.40 yrs), we used daily full-disk filtergrams from SDO/AIA 193 \AA. Table 1 presents a summary description of the original data.

Based on the filtergrams, we plotted synoptic maps of the EUV emission and detected low-emission patterns which correspond to CHs. The sequence of operations was the following. Using a square root transformation, we pre-processed the maps to allow adequate extraction of low-intensity features. Then, we applied the watershed segmentation, the standard \textit{k}-means clustering, and the fuzzy \textit{C}-means algorithm to detect CHs \citep{Reiss15,Caplan16}. The clustering algorithm estimates a membership function that quantifies pixel intensities to separate bright and dark features in a gray-scale image. For practical implementation of the techniques, we used codes developed by \cite{Bonnet02} and \cite{Caplan16}. The segmentation and clustering algorithms gave reasonable results which are similar throughout the solar cycles. Performing the post processing, we had to remove the remaining parts of filament channels. In order to correct our CH maps properly, we compared them with the daily CH charts from the Space Weather Prediction Center (\url{www.ngdc.noaa.gov/stp/space-weather/solar-data/solar-imagery/composites/full-sun-drawings/boulder}).

All the maps are presented in FITS format. They are in the cylindrical equal-area projection and have 360 points in longitude and 180 points in sine latitude with equidistant steps on the axes. The CH maps are coded with $+$1 and $-$1 according to their dominant polarities against the zero valued background.

There are gaps in SOLIS/VSM observations for CRs 2152\,--\,2155, 2163, 2166, 2167 and 2176 (2014.49\,--\,2014.79, 2015.31\,--\,2015.39, 2015.54\,--\,2015.68, and 2016.28\,--\,2016.36 yrs). We filled them with appropriate NSO/GONG data. 
To adjust the substituted data for each gap, we evaluated adjustment coefficients from pixel-to-pixel comparison of SOLIS/VSM and NSO/GONG maps for four CRs: two preceding a gap and two succeeding it. In each case,  we calculated the parameters of the linear regression equation, using the method of the reduced major axis \citep{Davis86}.

\begin{table}    
\setlength{\tabcolsep}{5pt}  
\caption{ Data sources. 
}
\label{T-simple}
\begin{tabular}{lllll} 
\hline 
Instru-& Spectral line,      & CRs \ \ \ & \ Data archive location \ \ \        & Reference        \ \   \\
ment   &  {\AA}              &     \ \ \ &   \ \ \       &                  \ \   \\
\hline 
      &                      &         &\underline{Synoptic magnetic maps:}              \\
NSO/  & Fe~{\sc i} λ8688  &1909-2007&\url{ftp://vso.nso.edu/kpvt/}    &\cite{Jones92}          \\
KPVT  &                      &         &\url{synoptic/mag}              &                        \\
SOLIS/&Fe~{\sc i} λ6302.5&2007-2189&\url{solis.nso.edu/0/}   &\cite{Balasubramaniam11}\\
VSM   &                      &         &\url{vsm/crmaps}                &                        \\
NSO/  &Ni~{\sc i} λ6768  &2047-2190&\url{ftp://gong2.nso.edu/}       &\cite{Harvey96}         \\
GONG  &                      &         &\url{QR/mqs}                    &\cite{Leibacher99}      \\
\hline 
      &                      &         &\underline{Synoptic CH maps:}                    \\
NSO/  &He~{\sc i} λ10830 &1626-2003&\url{ftp://nsokp.nso.edu/}&\cite{Jones92}          \\
KPVT  &                      &         &\url{kpvt/synoptic}     &                        \\
SoHO/ &Fe~{\sc xii} λ195 &1909-2123&\url{lasco-www.nrl.navy.mil}&\cite{Delaboudini95}    \\
EIT   &                      &         &     &                        \\
SDO/  &Fe~{\sc xii} λ193 &2124-2163&\url{secchi.nrl.navy.mil}&\cite{Lemen12}          \\
AIA   &                      &         &     &                        \\
\hline 
      &                      &         &\underline{Daily corona images:}                 \\
SDO/  &Fe~{\sc xii} λ193 &2164-2190&\url{www.solarmonitor.org}       &\cite{Lemen12}          \\
AIA   &                      &         &                                 &                        \\
\hline 
\end{tabular}
\end{table}

\section{Method}
\label{S3-Mthd}
Both series of maps were averaged over 5 CRs. A map for middle Carrington rotation (MCR) remained without changes. Data of two previous and two following maps were differentially rotated to MCR using the laws:
\begin{equation}		\label{Eq-Om1}
{\mathrm{\omega_{MF} = 14.33(\pm 0.054) - 2.12(\pm 0.35) sin^2 \varphi - 1.83(\pm 0.38) sin^4 \varphi}},
\end{equation}
\begin{equation}     \label{Eq-Om2}
{\mathrm{\omega_{CH} = 14.23(\pm 0.03) - 0.40(\pm 0.10) sin^2 \varphi}}.
\end{equation}

Here, $\varphi$ is the heliographic latitude, $\omega_{\mathrm {MF}}$ and $\omega_{\mathrm {CH}}$ are the angular velocities expressed in deg day$^{\mathrm{-1}}$ for magnetic fields \citep{Howard90} and CHs \citep{Timothy75}, respectively.

The averaged distributions of magnetic field were denoised using a wavelet based technique. Then, we detected the regions of long-lived ACs (areas where magnetic-field modulus was not less than 10 G) and boundaries of unipolar magnetic regions (contours ±1 G) on the maps. The detected magnetic features were superposed on the averaged CH maps.

The averaged CH maps were denoised using a wavelet decomposition technique. So, they are presented as distributions in a continuous set of values from −1 to +1. We avoid a discrete set of values in increments of 0.2 that originates in averaging of the maps with the three-level encoding. The smoothed averaged maps demonstrate well-defined ECHs as domains of existence of long-lived CHs.

The averaged maps of magnetic fields reveal long-lived ACs and associated UMRs. The averaged CH-maps show the regions of frequent CH-appearance, taking their dominant polarities into account. At those locations, relatively short-lived CHs tend to form long-lived structures \,-- ECHs. Thus, long-lived ECHs become visible near decaying ACs in the averaged maps.

\section{Representation of Results}
\label{S4-ResRepres}
The method of synoptic map assimilation shows clearly the interrelation of evolution of long-lived features in magnetic fields and in CHs, which is essential to understand PCH formation. It enables us to study the prehistory of PCH origin in detail: the development of long-lived ACs, the poleward migration of the UMRs  associated with the AC decay, and the formation of high-latitudinal ECHs. In this section, we demonstrate the reliability of the results obtained using this method with the example of the sequence of events in CRs 2149\,--\,2153 (2014.27\,--\,2014.64 yrs). 

During 2014, many large ARs and long-lived ACs were observed. The highest concentration of magnetic flux took place in the southern hemisphere at longitudes 200$^\circ$\,--\,300$^\circ$. Here, upon decay of large ACs, some extensive UMRs appeared and a surge of negative (trailing) polarity has formed. Later, the surge reached the South Pole. So, the conditions for PCH formation were prepared.

\begin{figure}    
\centerline{\includegraphics[width=0.9\textwidth,clip=]{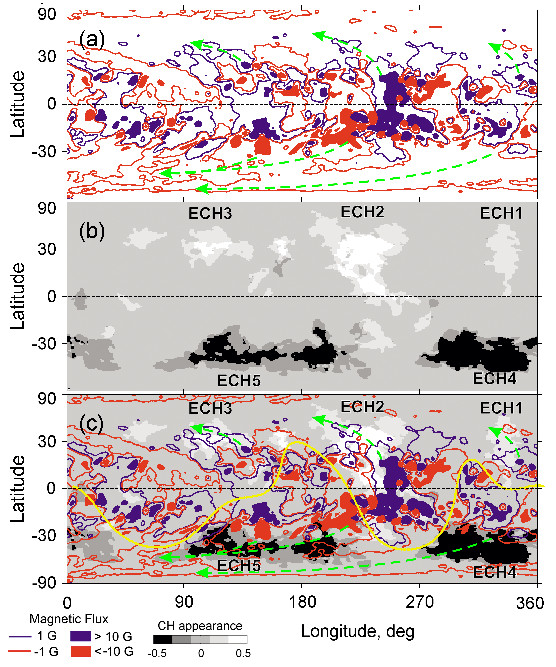} 
              }  
  \caption{Averaged distributions of magnetic fields (a), CHs (b), and their superposition (c) for MCR 2151. \textit{Blue contours} correspond to magnetic field of 1~G, and \textit{red contours} to -1~G. The fields greater than 10~G are in \textit{blue}, and the fields less than -10~G are in \textit{red}. \textit{Green arrows} point out the magnetic surges. The CH appearance is in \textit{black-to-white} and accords to  dominant magnetic field polarities.  The neutral line is shown in \textit{yellow}.}    
               
   \label{F-represMCR2151}
   \end{figure}

Figure~\ref{F-represMCR2151}a shows the averaged distribution of magnetic fields differentially rotated to MCR 2151 (2014.41\,--\,2014.49 yrs). Here, we use 10 G in modulus as a threshold value at the edges of decaying ARs. Red and blue contours approximately correspond to the boundaries of UMRs, which formed during decay of the long-lived ACs and which combined the negative and positive polarities of magnetic field. Arrows depict the surges shaped due to the meridional transport and differential shearing of remnant magnetic-fields. Clearly, according to the differential rotation profile, the surges at mid-latitudes are inclined equatorward at an angle which is greater than those observed at high latitudes. Previously, \cite{Golubeva16} noted that the high-latitudinal surges of negative polarity formed due to the decay of pre-existing ACs, then formed a ring structure around the South Pole during their meridional transport and finally caused the change of dominant polarity at the South Pole.

Figure~\ref{F-represMCR2151}b shows the averaged CH distribution for MCR 2151. Light and dark gray-tones mark the domains of frequent CH appearance of positive and negative polarity, according to the colorbar. In the northern hemisphere, we labeled three regions of the most likely appearance of positive polarity CHs: ECH1, ECH2, and ECH3. In the southern hemisphere, two regions of frequent appearance of negative polarity CHs can be seen: ECH4 and ECH5. 

Comparison of Figures~\ref{F-represMCR2151}a and \ref{F-represMCR2151}b demonstrates that ECH4 and ECH5 formed in the latitudinal belt of negative polarity UMRs. It is also clear that the ECHs  usually formed near the major long-lived ACs, or the nests of activity \citep{Golubeva16}.

Figure~\ref{F-represMCR2151}c presents the superposition of distributions given in Figures~\ref{F-represMCR2151}a and \ref{F-represMCR2151}b, and the neutral-line position on the source surface at 2.5 solar radius for CR 2151. The coronal magnetic field was calculated in the potential approximation with classical boundary conditions \citep{Petrie13}. It is characterized by a four-sector structure. We can see interconsistency between all the elements given in the final complex map. A series of such complex maps for MCRs 1911\,--\,2187 (1996.49\,--\,2017.18 yrs) has been plotted and analyzed. The results of analysis are presented in further sections.

\begin{figure}    
   \centerline{\includegraphics[width=0.75\textwidth,clip=]{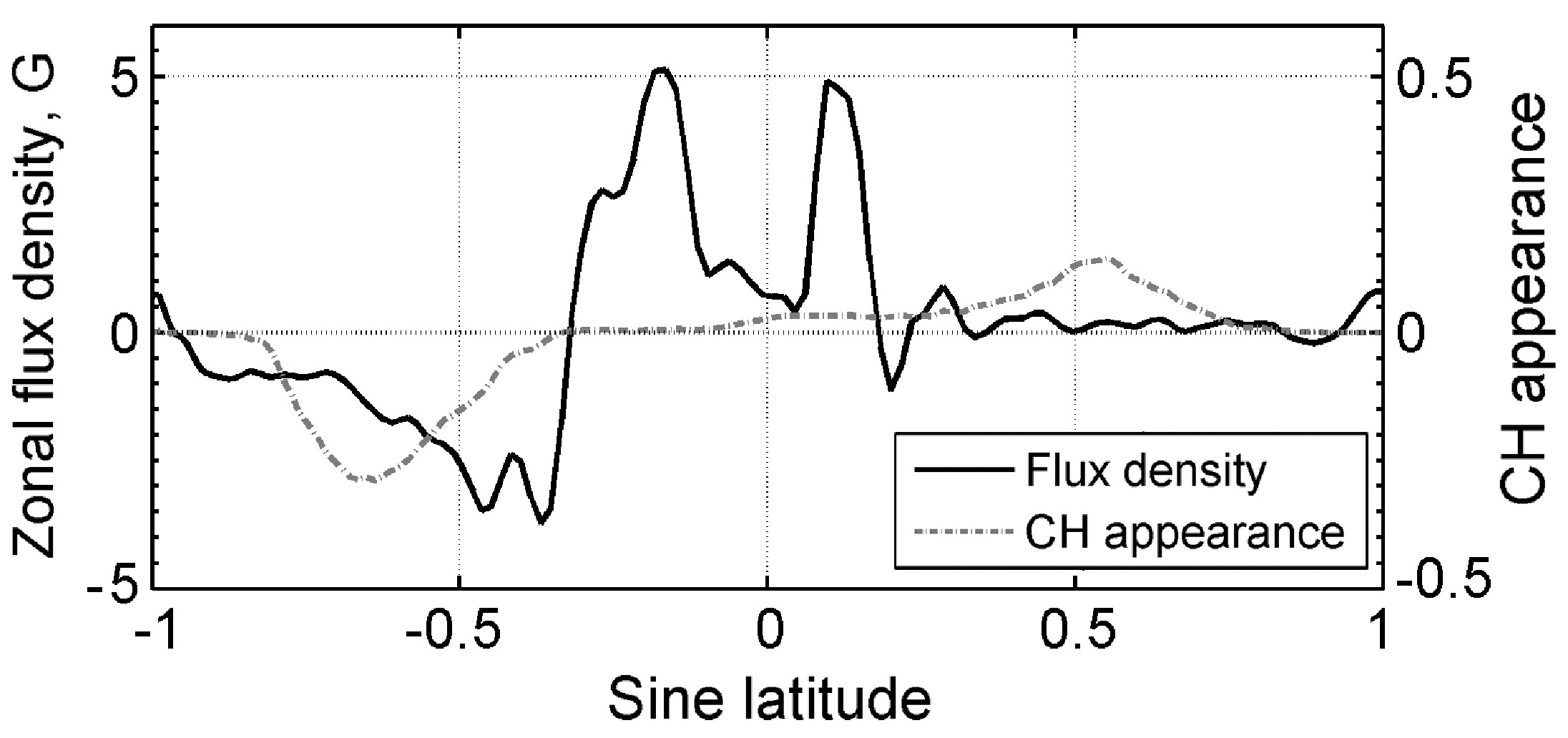} 
              }
 \caption{Latitudinal distributions of  magnetic fields and CHs, which are longitude averaged, for MCR 2151.} 
 
   \label{F-longaver2151}
   \end{figure}

Figure~\ref{F-longaver2151} shows longitude averages of magnetic flux density and of CH appearance for MCR 2151. The shown profiles describe zonal structure in the distributions. 
Relatively strong fields are concentrated at low latitudes from $-15^\circ$ to 15$^\circ$.
In the southern hemisphere, the magnetic-field polarities had already been separated by the time considered. The negative polarity dominated in the latitudinal interval $-90^\circ$\,--\,$-20^\circ$, and the positive polarity dominated closer to the equator. In the middle latitudes of the northern hemisphere, we can observe a poorly-recognizable predominance of positive polarity. In the CH distribution averaged over longitudes, the regions of the frequent appearance of negative polarity CHs are in the longitudinal interval $-50^\circ$\,--\,$-30^\circ$, but the positive polarity CHs most often appear in the longitudinal interval 25$^\circ$\,--\,45$^\circ$. This zonal structure in CH distribution appears due to the formation of ECH belts at mid latitudes.

\section{Evolution of Magnetic Field and Coronal Holes in Cycles 23 and 24}
\label{S5-twoCycles}

We consider features of the current 11-year cycle as compared with the previous Cycle 23. Such an analysis enables us to examine solar activity evolution during the complete magnetic cycle. 

\begin{figure}    
   \centerline{\includegraphics[width=0.95\textwidth,clip=]{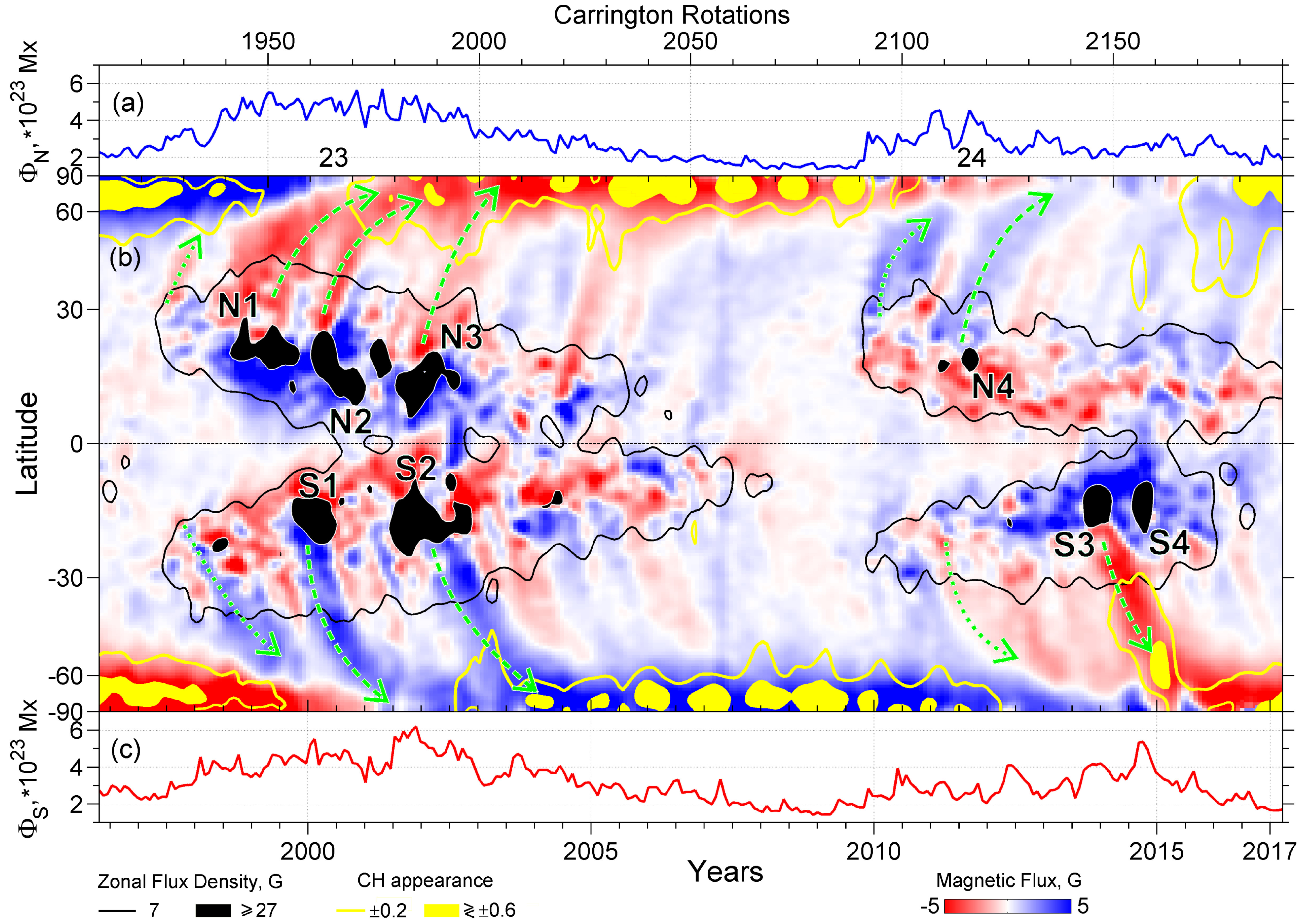} 
              }
         \caption{Time-latitude analysis of magnetic flux and CHs in Cycles 23 and 24 (numbered in the \textit{top panel}). Variations of absolute magnetic flux in the northern ($\Phi_{\mathrm {N}}$) and southern ($\Phi_{\mathrm {S}}$) hemispheres are shown in the \textit{top} (a) and \textit{bottom} (c) \textit{panels}, respectively. The time-latitude diagram is shown in \textit{panel} b. The longitude-averaged magnetic fields are in \textit{red-to-blue}. \textit{Black contours} of 7~G in modulus depict the boundaries of sunspot activity zones. Domains of high zonal flux density above 27~G in modulus are in \textit{black}. \textit{Green arrows} point out the remnant flux surges. Domains of CH appearance at a level above 0.6 in modulus are in  \textit{yellow}. \textit{Yellow contours} correspond to CH appearance $\pm$0.2.}

   \label{F-diagram}
   \end{figure}

To study the evolution of photospheric magnetic fields, we analyzed original synoptic maps from NSO/KPVT and SOLIS/VSM for the period 1996\,--\,2017.
Figures~\ref{F-diagram}a and \ref{F-diagram}c show the changes of absolute magnetic flux in the northern and southern hemispheres, respectively.
We performed a time-latitude analysis of the maps. 
Here, each map was averaged over longitude to obtain a latitudinal profile characterizing the zonal structure of magnetic field for every CR. 
Then, we arranged the obtained profiles in chronological order for CRs 2049\,--\,2190.  
So, the central panel (3b) shows the time-latitude diagram that is composed of zonally-averaged  synoptic maps. The distribution was denoised, using the wavelet decomposition technique. 
The diagram vividly demonstrates evolution of solar magnetic fields and their cyclic rearrangements. 
It also shows cause-effect relations between zones of intense sunspot activity, remnant flux surges, and long-term patterns in CH appearance.
 
Analysis of the results given in Figure~\ref{F-diagram} reveals the following picture.

During the 11-year cycles, sunspot activity zones migrated equatorward according to Spoerer's law (Figure~\ref{F-diagram}b). At the beginning of each cycle, small ARs emerged at latitudes about 30$^\circ$. After their decay, trailing-polarity UMRs formed near the high-latitudinal boundaries of the sunspot zones. In Cycles 23\,--\,24, the first surges (shown with dotted arrows) were associated with emergence of relatively small magnetic fluxes. As the surges approached the poles, the cancellation of opposite magnetic polarities led to a decrease in the latitudinal extent of the polar caps.

Highly-concentrated magnetic-fluxes (greater than 27~G in modulus) were observed during the epochs of high solar-activity. They are shown in black (Figure~\ref{F-diagram}b). The largest of them are denoted as N1, N2, N3 and N4 in the northern, and as S1, S2, S3 and S4 in the southern hemisphere. They depict the latitudes and timing of long-lived ACs. These fluxes caused the trailing-polarity surges.

For example, regions N2 and S1 emerged at latitudes 15$^\circ$\,--\,25$^\circ$ approximately at the same time in 2000\,--\,2001, when the primary maximum of Cycle 23 was observed. As the ACs evolved and after their decay, the remnant magnetic fields dissipated and formed UMRs. Diffusion and advection of the fields determined the further evolution of the UMRs. Due to meridional flows, the UMRs became the surges. The most extensive surges reached the polar zones. They are marked with dashed arrows.

The secondary maximum of activity in Cycle 23 was associated with the emergence of N3 and S2. These highly-concentrated magnetic fluxes also led to formation of trailing-polarity surges, that approached to the poles and increased the latitudinal extent of the polar UMRs. The remnant magnetic fluxes were transported from the sunspot-activity zones to the poles during about 1.5\,--\,2 years.

In the current cycle, the level of magnetic activity was significantly lower than the one observed in Cycle 23 (Figures~\ref{F-diagram}a and \ref{F-diagram}c). In addition, significant asymmetry and asynchrony were obvious in the activity development of both hemispheres (Figure~\ref{F-diagram}b). In the second half of 2011, there were long-lived ACs in the northern hemisphere. After they decayed, remnant magnetic fluxes formed extensive UMRs of positive (trailing) polarity. Further, meridional flows transported these weak magnetic fields to the northern polar zone and caused the change of dominant polarity from negative to positive. Transport of remnant magnetic fluxes from region N4 to the Pole took about 1.6 years \citep{Mordvinov14, Sun15}.

The peak of activity in the southern hemisphere took place in 2014. Its level was much higher than the one observed in the northern hemisphere in 2011 (Figures~\ref{F-diagram}a and \ref{F-diagram}c). Regions S3 and S4 in the time-latitude diagram (Figure~\ref{F-diagram}b) characterize the relatively-high concentration of the emergent magnetic-flux in the southern hemisphere. Decay of the ACs, that were observed in 2014, caused formation of extended remnant-flux surge, which changed dominant polarity at the south pole. However, this regular picture was disturbed by the leading-polarity surges which arose after the decay of ARs characterized by non-Joy's law tilts \citep{Yeates15, Mordvinov15,Mordvinov16}.

Our preliminary analysis of the averaged distributions for CR 2151 have shown causal relations between the zonal structure of open magnetic-fields and the remnant fields (Figure~\ref{F-represMCR2151}). Therefore, we applied the time-latitude analysis to the sequence of CH synoptic maps for CRs 1910\,--\,2190. In Figure~\ref{F-diagram}b, yellow contours outline the domains of frequent CH appearance. CH appearance in these domains exceeds 0.2. They encompassed the solar polar caps, when stable PCHs existed there. Yellow spots correspond to CH appearance equal to or above 0.6 in modulus.

In the southern hemisphere, a wide domain of frequent CH appearance occurred in 2014\,--\,2015. The mid- and high-latitudinal CHs have formed after the decay of multiple ACs including S3 and S4 during the period of high magnetic activity. Such a CH evolution could be interpreted as a systematic migration of CHs \citep{Harvey02}. The stable PCH formed at the South Pole in early 2015 and evolved till now (\textbf{October} 2017). 
Figure~\ref{F-diagram}b shows variations in size and visibility of the PCH. Such annual changes are typical for PCHs.

At the North Pole, the PCH formed later despite the fact that the dominant polarity there changed in early 2013. In the northern hemisphere, magnetic activity was very low through 2013\,--\,2015. 
Nevertheless, a huge domain of frequent CH appearance formed near the North Pole. This domain is outlined with yellow contour.
The yellow spot shows the north PCH that has persisted since mid-2016.

\begin{figure}    
   \centerline{\includegraphics[width=0.75\textwidth,clip=]{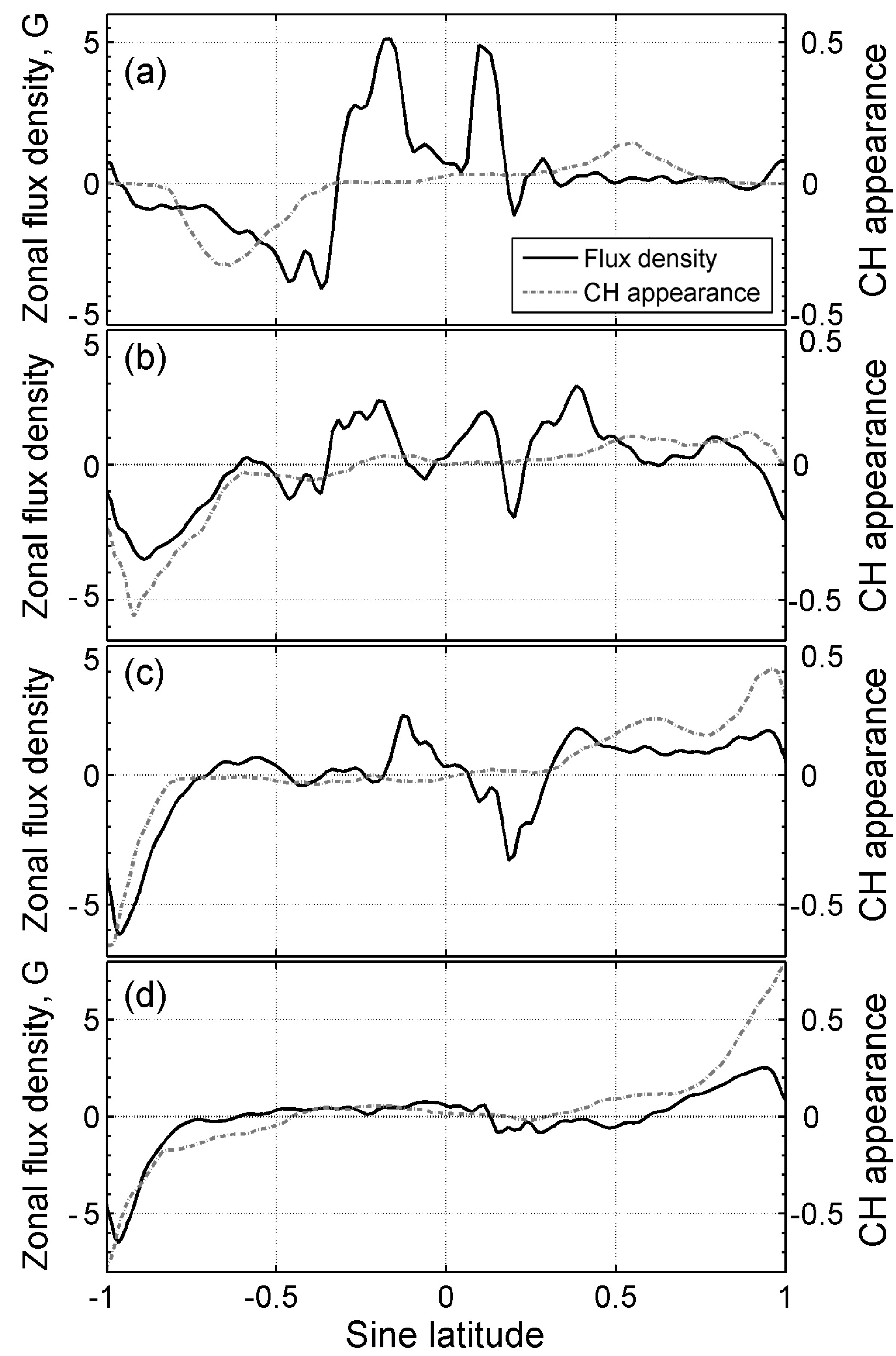} 
              }
  \caption{Latitudinal distributions of  magnetic fields and CHs, which are longitude averaged, for CRs 2159 (a), 2164 (b), 2177 (c), and 2186 (d).}  

   \label{F-longaver4CRs}
   \end{figure}

	It is of interest to compare conditions of CHs formation in the solar hemispheres. 
Figure~\ref{F-longaver4CRs} shows both magnetic flux density and CH appearance, which were zonally averaged, for CRs 2159, 2164, 2177, and 2186 (2015.01\,--\,2015.09, 2015.39\,--\,2015.46, 2016.36\,--\,2016.43, and 2017.03\,--\,2017.10 yrs). These distributions include causally related features in zonal structure, by analogy with Figure~\ref{F-longaver2151}.

In the southern hemisphere, high magnetic activity at latitudes less than 20$^\circ$ resulted in significant poleward remnant-flux of negative (trailing) polarity (Figures~\ref{F-longaver4CRs}a and \ref{F-longaver4CRs}b). The averaged distributions of CH appearance resemble the averaged flux-density profiles. These conditions proved to be favorable for the  south PCH formation. It has been completed by CR 2177 (Figure~\ref{F-longaver4CRs}c). 
The further development of the north PCH demonstrated its strengthening and stability (Figure~\ref{F-longaver4CRs}d).
In the northern hemisphere, weak flux-densities at latitudes less than 20$^\circ$ resulted in weak remnant flux of predominantly positive (trailing) polarity at higher latitudes. 
High magnetic activity in the southern hemisphere resulted in a significant remnant flux of negative (trailing) polarity that was transported to high-latitudes. 
These conditions proved to be favorable for the fast formation of the south PCH.

Thus, the time-latitude analysis demonstrates cyclic transformations of the magnetic flux: the emergence of ARs, their further decay, and the resulting formation of remnant flux surges  which reach the polar zones. As the UMRs of trailing-polarities approached the poles, the high-latitudinal ECHs were accumulated to form the PCHs.

\section{Formation of Polar Coronal Holes in the Current Cycle} 
\label{S6-PCHform}

\begin{figure}    
   \centerline{\includegraphics[width=0.85\textwidth,clip=]{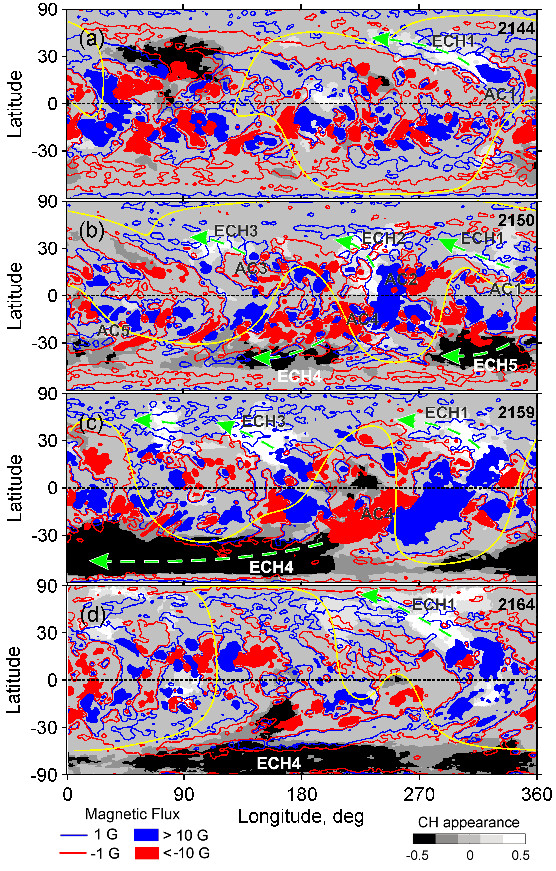} 
              }
   \caption{Main steps of the south PCH formation. The averaged maps for MCRs 2144 (a), 2150 (b), 2159 (c), and 2164 (d) show ECHs in \textit{black-to-white} according to the dominant magnetic field polarities  Here, the absolute CH appearance values exceeding 0.5 are set equal to $\pm$0.5 subject to their signs. \textit{Blue contours} correspond to magnetic field 1~G, and \textit{red contours} to -1~G. Magnetic fields greater than 10~G are in \textit{blue}, and the fields less than -10~G are in \textit{red}. \textit{Green arrows} point out the magnetic surges. The neutral line is shown in \textit{yellow}.}
   \label{F-southPCH}
   \end{figure}

\begin{figure}    
   \centerline{\includegraphics[width=0.85\textwidth,clip=]{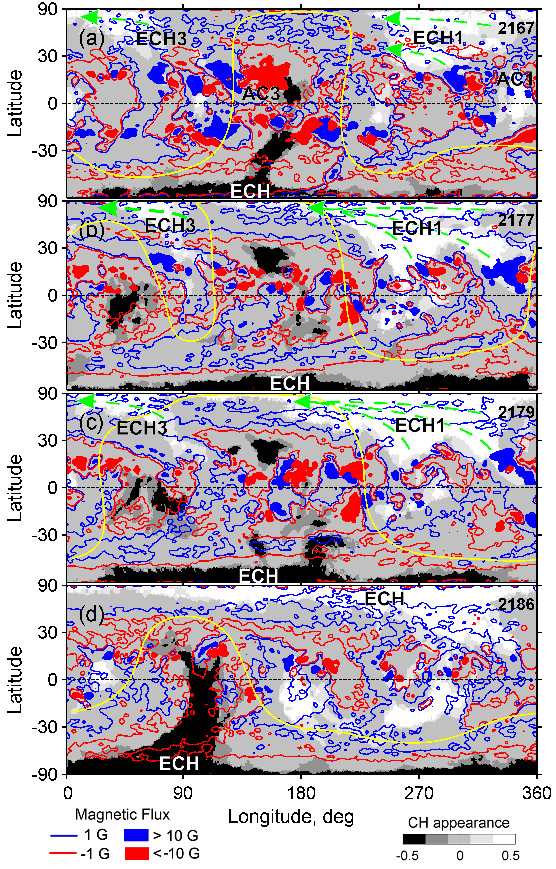} 
              }
   \caption{Main steps of the north PCH formation. The averaged maps for MCRs 2167 (a), 2177 (b), 2179 (c), and 2186 (d) show ECHs in \textit{black-to-white} according to the dominant magnetic field polarities.  Here, the absolute CH appearance values exceeding 0.5 are set equal to $\pm$0.5 subject to their signs. \textit{Blue contours} correspond to magnetic field 1~G, and \textit{red contours} to -1~G. Magnetic fields greater than 10~G are in \textit{blue}, and the fields less than -10~G are in \textit{red}. \textit{Green arrows} point out the magnetic surges. The neutral line is shown in \textit{yellow}.} 
   \label{F-northPCH}
   \end{figure}

Analyzing the complete set of composite averaged maps, we specified the main stages of the asynchronous PCH formation in the southern and northern hemispheres. Figures~\ref{F-southPCH} and \ref{F-northPCH} show compactly these long-duration processes for the south and north PCHs, relatively. Figure~\ref{F-southPCH} includes the maps for MCRs 2144, 2150, 2159, and 2164 (2013.89\,--\,2013.97, 2014.34\,--\,2014.42, 2015.01\,--\,2015.09,  and 2015.39\,--\,2015.46 yrs).  Figure~\ref{F-northPCH} includes the maps for MCRs 2167, 2177, 2179, and 2186 (2015.61\,--\,2016.68, 2016.36\,--\,2016.43, 2016.51\,--\,2016.58, and 2017.03\,--\,2017.10 yrs). Shown in gray scale according to their embedded polarities, the averaged CH distributions enable us to identify long-term features in CH appearance. Here,  black and white correspond to ECHs. They are the domains of frequent CH appearance. Dark and light halftones show additional details in ECH evolution against the mid-gray background. Red and blue areas correspond to negative and positive polarities with magnetic field greater than 10~G in modulus, respectively. Color contours and other notations are the same as in Figure~\ref{F-represMCR2151}. 

Formation of the south PCH was attended by several ACs and a considerably higher level of magnetic activity in the southern hemisphere than in the northern one. We concentrate on AC1, AC2, AC3, AC4, and AC5 in Figure~\ref{F-southPCH}. The trailing-polarity UMRs with the embedded ECHs originated from each of these ACs. Arrows show ECH1, ECH2, ECH3, ECH4, and ECH5 shaped due to the differential rotation and meridional flows. 

In MCR 2144 (Figure~\ref{F-southPCH}a), AC2 and AC4 were compact. The trailing-polarity part of A2 and the leading-polarity part of AC4 have integrated in MCR 2150 (Figure~\ref{F-southPCH}b). The structural peculiarities of evolving AC4 attract attention in MCR 2159 (Figure~\ref{F-southPCH}c). First, it is the largest of the ACs and represents a region of the greatest yield of magnetic flux at that time. Second, its region of leading (positive) polarity significantly exceeds its region of trailing polarity in area extent and evidently integrates with the region of AC2 trailing polarity in the northern hemisphere.

The high-latitudinal extended ECH4 of positive polarity, which stretched from AC4, began to form in MCRs 2145\,--\,2149. At the same time, there was a chain of weak ECHs of negative polarity at 130$^\circ$\,--\,180$^\circ$ longitudes. Then it merged with high-latitudinal ECH5 associated with AC5. At that, ECH4 enhanced and moved poleward. Thus, it was formed due to the accumulation and averaging of contributions from many CHs, which occupied increasingly-higher latitudes as the negative-polarity UMR was transported to the South Pole. 

No wonder that, in Figure~\ref{F-southPCH}c, the decaying AC4 is accompanied by extensive high-latitudinal ECH4, which occupies the broad ring-shaped UMR formed by the merger of the trailing (negative) polarities of closely-spaced ACs in the southern hemisphere. This UMR imposed the organizing effect on high-latitudinal CHs by forming ECH out of them. Some extensions directed toward other ARs indicate additional sources of ECH4. One could expect that further enhancement and poleward transport of ECH4 might cause gradual formation of the south PCH. However, the scenario was more dramatic, apparently because of the instability caused by AC4 decay.

By MCR 2159, a positive-polarity UMR has occupied the South Pole, spanning all longitudes, as is shown in Figure~\ref{F-southPCH}c. Here, we can note the initial eastward  stretch of the leading (positive) polarity UMR from AC4 to a longitude of 223$^\circ$. This leading polarity surge  moved to longitudes  201$^\circ$, 162$^\circ$, 128$^\circ$, and 131$^\circ$ successively in MCRs 2160\,--\,2163. In MCR 2164 it reached 95$^\circ$ (Figure~\ref{F-southPCH}c). Then the UMR
began to disintegrate due to cancellation of the opposite polarities. 

It is noteworthy that at the initial stage of positive polarity UMR stretching from AC4, the high-latitudinal negative polarity ring-shaped UMR occupied the South Pole across all longitudes (Figure~\ref{F-southPCH}c).  However, it took time for the associated ECH4 covering increasingly higher latitudes to become the south PCH in MCR 2164 (Figure~\ref{F-southPCH}d). At the early stage of its formation, the south PCH was very asymmetric and had a transequatorial extension along ~155$^\circ$ longitude. It is notable that the extension was resting on the negative polarity regions of two large decaying ACs, over the trailing polarity region in the southern, and over the leading polarity region in the northern hemisphere. It was practically stable up to MCR 2170. It should be noted that the predominant-polarity at the South Pole
was uncertain in MCRs 2165\,--\,2171, but the polarity has remained negative since MCR 2172.
          
The process of north PCH formation was different. Long-lived AC1, AC2, and AC3 (Figure~\ref{F-southPCH}) played an important role. The decay of AC2, AC3, and adjacent ARs resulted in positive-polarity UMRs and further formation of high-latitude surges approaching the North Pole. In parallel with the remnant flux transport, ECH2 and ECH3 appeared within these surges (Figures~\ref{F-southPCH}b and \ref{F-southPCH}c). The subpolar ECH3 was possibly associated with the low-activity impulse N4 in 2011\,--\,2012 (Figures~\ref{F-diagram}a and \ref{F-diagram}b). ECH2 disappeared after MCR 2163. By MCR 2164 (Figure~\ref{F-southPCH}d), ECH1 and ECH3 had propagated to higher latitudes and occupied a wider longitudinal interval. During this transformation, open magnetic flux covered most of the northern polar zone (Figure~\ref{F-southPCH}d).
At that time, negative polarity was dominating at the North Pole, and positive polarity UMRs approached from lower latitudes. 

In MCR 2167, ECH1 and ECH3 reached the North Pole due to meridional transport and merged into a connected structure, which  occupied about  half of the longitudes at the highest latitudes (Figure~\ref{F-northPCH}a). Its trans-equatorial extension began to form at longitudes about ~230-280$^\circ$ (Figures~\ref{F-northPCH}a and \ref{F-northPCH}b). In MCR 2179 (Figure~\ref{F-northPCH}c), the ECH structure turned into the north PCH that covers almost all longitudes. At this stage of its existence, the north PCH looked highly asymmetric. It had three extensions, one of which was trans-equatorial. This extension existed and evolved,  until October 2017. At the North Pole, the magnetic polarity was mostly negative in MCRs 2161\,--\,2165. From MCR 2168, it was mostly positive, and from MCR 2179  (Figure~\ref{F-northPCH}d) it is exceptionally positive. 

The evolution of ACs, UMRs, and ECHs  led to further significant changes in the sizes and shapes of the PCHs.  Tracing the evolution of the latter, we have noticed the following details.

A weak equator-ward extension of the south PCH formed at longitudes around 80$^\circ$ in MCR 2167. It reached mid-latitudes in MCR 2170, crossed the equator in MCR 2173 and decayed into two parts in MCR 2177 (Figure~\ref{F-northPCH}b). Obviously, this decay was caused by the mid-latitudinal UMR-strip of positive polarity which approached from the west. This formed the sub-equatorial ECH moving westward in Carrington system due to the differential rotation. The south PCH had both a trans-equatorial extension at longitude 180$^\circ$ and a newly-emerged equator-ward extension at longitude 145$^\circ$ in MCR 2179  (Figure~\ref{F-northPCH}c). The first one had decreased to zero by MCR 2186  (Figuree~\ref{F-northPCH}d). The second one had merged with the sub-equatorial ECH, having formed the extended trans-equatorial extension of the south PCH, which stretched to ACs of the northern hemisphere (Figure~\ref{F-northPCH}d).

The eastern-most extension of the north PCH disappeared, and a sub-equatorial positive-polarity ECH formed at longitudes around 170$^\circ$ in MCR 2183. By MCR 2186 (Figure~\ref{F-northPCH}d), the ECH joined with the two previously-merged western-extensions of the north PCH. Figure~\ref{F-northPCH}d shows the very long total extension crossing the equator twice. It was associated exclusively with the sub-equatorial ACs of the northern hemisphere, which were spaced at about 100$^\circ$ in longitude. It reached high latitudes of the southern hemisphere due to the configuration of the underlying UMRs. Note, it was not associated with ACs of the southern hemisphere.

It is remarkable that when the ACs emerged on the Sun, the equatorward extensions stretched toward them from PCHs. Thus, the extension of the south PCH has formed by MCR 2186 (Figure~\ref{F-northPCH}d), and stretched to the northern-hemisphere AC. This trans-equatorial extension induced the recurrent geomagnetic-storms during several CRs in the descending phase of the current cycle. This was the time of high geomagnetic-activity. Obviously, the PCH extensions, as well as the mid- and low-latitudinal ECHs substantially determined the neutral-line configuration. Hence, they essentially specifyed the sector structure of the interplanetary magnetic-field. Thus, the positions of low-latitudinal ECHs in MCRs 2164 (Figure~\ref{F-southPCH}d) and 2177 (Figure~\ref{F-northPCH}b) corresponded to the four-sector structure of the magnetic field. Subsequent changes in the ECH configuration resulted in the two-sector structure observed in MCR 2186 (Figure~\ref{F-northPCH}d).

\begin{figure}    
   \centerline{\includegraphics[width=0.9\textwidth,clip=]{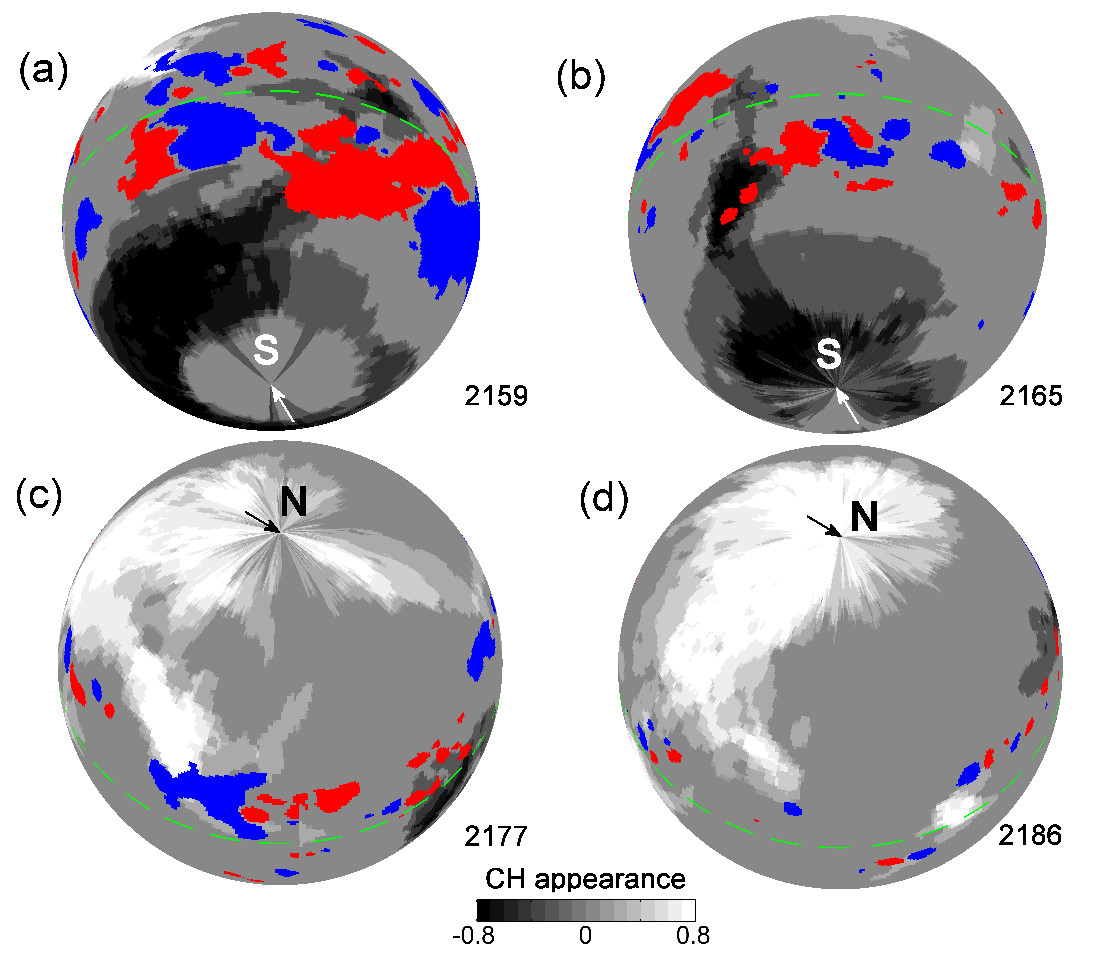} 
              }
    \caption{The polar views derived from 3D-reconstruction of the averaged maps: the South  Pole (S) for MCRs 2159 and 2165 (a and b), and the North Pole (N) for MCRs 2077 and 2186 (c and d). The CH appearance saturated at $\pm$0.8 is shown in \textit{black-to-white}, according to the magnetic field polarities. Fields greater than 10~G are in \textit{blue}, and fields less than -10~G are in \textit{red}. The poles are marked with \textit{arrows} directed along the zero meridian. The equator line is \textit{green}.}
   \label{F-poles}
   \end{figure}

Figure~\ref{F-poles} shows the Sun$'$s polar views derived from the 3D reconstruction of the averaged maps for MCRs 2159 and 2165 (a,b) around the South Pole and for MCRs 2177 and 2186 (c,d) around the North Pole. The South and North Poles are indicated with S and N arrows directed along the zero meridian. The south circumpolar ECH formed in its relation to the high-latitudinal surge and UMRs which were shaped after the decay of the giant AC4 in 2014. After two CRs, this ECH became the south PCH with the trans-equatorial extension passing along boundaries of ACs situated in the both hemispheres (Figure~\ref{F-poles}b). At the early stage of its formation, the north PCH demonstrated relationships with UMRs formed during decay of separate ARs. Notably, these relationships defined the initial PCH structure, which remained for a long time.

The evolution of ECHs and magnetic flux is shown in a movie that is presented in the Electronic Supplementary Material. This movie shows the composite synoptic maps in the same notations. It illustrates the formation of the PCHs and the open-flux rearrangements in more detail.

\section{Conclusions}
\label{S7-Conc}
 To study the global reorganization of open magnetic flux and the formation of PCHs in the current cycle,  we developed a method of assimilation of synoptic maps of magnetic fields and CHs. The present approach is useful to understand CH evolution and the general regularities in global changes of magnetic fields during an activity cycle.
 
It has been established that the defining role in the cyclic reorganization of open magnetic-flux belongs to the ECHs, which form in the UMRs associated with the decaying ACs. The ECHs interact with the ARs and ACs due to the reconnection of open and closed magnetic fields. In the global reorganization of magnetic fields, the ECHs merge together. At that, they form extensive trans-equatorial structures, through which the magnetic flux passes from one hemisphere to another. As UMRs move toward the poles, the high-latitudinal ECHs formed. Gradual accumulation and merging of ECHs in the polar caps cause formation of the PCHs. Thus, the peculiarities of PCH formation in the current cycle can be explained by the cause-effect relations between the ACs, UMRs, and ECHs.

In particular, the decay of large ARs and long-lived ACs observed in 2014 resulted in the appearance of  huge UMRs of negative polarity. The highly-concentrated remnant-flux led to 
the fast cancellation of the old polarity. In parallel with the poleward transport of trailing-polarity UMRs, the high-latitudinal ECHs restructed and became the southern PCH.
In the northern polar-zone, the alteration of UMRs of opposite polarities led to the intricate configuration of magnetic fields. Therefore, it took much longer for accumulation of remnant-flux sufficient to build up uniform field in the north polar zone. 
These conditions led to the significant delay in formation of the stable north PCH.

The causality between the sunspot activity, the large-scale magnetic-fields, and the CH appearance was also demonstrated in their time-latitude behaviour.
 The ECH formation occurs in close connection with the evolution of ACs and further poleward transport of remnant fluxes. The step-by-step accumulation and merging of high-latitudinal ECHs results in formation of PCHs. 
Thus, both the specific spatio-temporal organization of the emergent strong magnetic-fields and the different conditions for the remnant-flux transport in the solar hemispheres caused the asynchronous PCH formation in the current cycle.

%

\begin{acks}

NSO/Kitt Peak data used here are produced cooperatively by NSF/NOAO, NASA/GSFC, and NOAA/SEL. 
The NSO/KPVT coronal hole data used here were compiled by K. Harvey and F. Recely using observations under a grant from the NSF.
This work utilizes SOLIS data obtained by the NSO Integrated Synoptic Program (NISP), managed by the NSO, which is operated by the Association of Universities for Research in Astronomy (AURA), Inc. under a cooperative agreement with the National Science Foundation.
This work utilizes data obtained by the Global Oscillation Network
Group (GONG) program, managed by the NSO. The data were acquired by instruments operated by the Big Bear Solar Observatory, High Altitude Observatory, Learmonth Solar Observatory, Udaipur Solar Observatory, Instituto de Astrof\'{\i}sica de Canarias, and Cerro Tololo Interamerican Observatory.
The authors also employed synoptic maps of the solar EUV emission prepared by SOHO/EIT science team and NASA/SDO and the AIA, EVE, and HMI science teams.
The authors are grateful to Svetlana Philippova for assistance in preparation of the English version and to the referee for the useful comments.

This work was supported by the Russian Foundation for Basic Research (project 17-02-00016) and by the SB RAS Program II.16.3.1.


\end{acks}

\section{Disclosure of Potential Conflicts of Interest}
The authors declare that they have no conflicts of interest.



\bibliographystyle{spr-mp-sola}
\bibliography{GolubevaMordvinov}  

\IfFileExists{\jobname.bbl}{} {\typeout{}
\typeout{****************************************************}
\typeout{****************************************************}
\typeout{** Please run "bibtex \jobname" to obtain} \typeout{**
the bibliography and then re-run LaTeX} \typeout{** twice to fix
the references !}
\typeout{****************************************************}
\typeout{****************************************************}
\typeout{}}

\end{article} 

\end{document}